\documentclass[aps,prc,showpacs,fleqn,floatfix,twocolumn,superscriptaddress,twoside]{revtex4}

\usepackage{graphicx} 
\usepackage{amsmath,amssymb,amsfonts}

\newcommand{\mean}[1]{\left\langle #1 \right\rangle}

\newcommand{ \la }{\langle}
\newcommand{ \ra }{\rangle}

\begin{document}


\title{Elliptic and hexadecapole flow of charged hadrons \\
in Au+Au collisions at $\sqrt{s_{NN}}$ = 200 GeV
}

\newcommand{\abilene}{Abilene Christian University, Abilene, Texas 79699, USA}
\newcommand{\banaras}{Department of Physics, Banaras Hindu University, Varanasi 221005, India}
\newcommand{\barc}{Bhabha Atomic Research Centre, Bombay 400 085, India}
\newcommand{\bnlcoll}{Collider-Accelerator Department, Brookhaven National Laboratory, Upton, New York 11973-5000, USA}
\newcommand{\bnlphys}{Physics Department, Brookhaven National Laboratory, Upton, New York 11973-5000, USA}
\newcommand{\caucr}{University of California - Riverside, Riverside, California 92521, USA}
\newcommand{\charlesczech}{Charles University, Ovocn\'{y} trh 5, Praha 1, 116 36, Prague, Czech Republic}
\newcommand{\chonbuk}{Chonbuk National University, Jeonju, 561-756, Korea}
\newcommand{\ciae}{China Institute of Atomic Energy (CIAE), Beijing, People's Republic of China}
\newcommand{\cns}{Center for Nuclear Study, Graduate School of Science, University of Tokyo, 7-3-1 Hongo, Bunkyo, Tokyo 113-0033, Japan}
\newcommand{\colorado}{University of Colorado, Boulder, Colorado 80309, USA}
\newcommand{\columbia}{Columbia University, New York, New York 10027 and Nevis Laboratories, Irvington, New York 10533, USA}
\newcommand{\czechtech}{Czech Technical University, Zikova 4, 166 36 Prague 6, Czech Republic}
\newcommand{\dapnia}{Dapnia, CEA Saclay, F-91191, Gif-sur-Yvette, France}
\newcommand{\debrecen}{Debrecen University, H-4010 Debrecen, Egyetem t{\'e}r 1, Hungary}
\newcommand{\elte}{ELTE, E{\"o}tv{\"o}s Lor{\'a}nd University, H - 1117 Budapest, P{\'a}zm{\'a}ny P. s. 1/A, Hungary}
\newcommand{\ewha}{Ewha Womans University, Seoul 120-750, Korea}
\newcommand{\fit}{Florida Institute of Technology, Melbourne, Florida 32901, USA}
\newcommand{\fsu}{Florida State University, Tallahassee, Florida 32306, USA}
\newcommand{\gsu}{Georgia State University, Atlanta, Georgia 30303, USA}
\newcommand{\hiroshima}{Hiroshima University, Kagamiyama, Higashi-Hiroshima 739-8526, Japan}
\newcommand{\ihepprot}{IHEP Protvino, State Research Center of Russian Federation, Institute for High Energy Physics, Protvino, 142281, Russia}
\newcommand{\illuiuc}{University of Illinois at Urbana-Champaign, Urbana, Illinois 61801, USA}
\newcommand{\instpasczech}{Institute of Physics, Academy of Sciences of the Czech Republic, Na Slovance 2, 182 21 Prague 8, Czech Republic}
\newcommand{\isu}{Iowa State University, Ames, Iowa 50011, USA}
\newcommand{\jinrdubna}{Joint Institute for Nuclear Research, 141980 Dubna, Moscow Region, Russia}
\newcommand{\jyvaskyla}{Helsinki Institute of Physics and University of Jyv{\"a}skyl{\"a}, P.O.Box 35, FI-40014 Jyv{\"a}skyl{\"a}, Finland}
\newcommand{\kek}{KEK, High Energy Accelerator Research Organization, Tsukuba, Ibaraki 305-0801, Japan}
\newcommand{\kfki}{KFKI Research Institute for Particle and Nuclear Physics of the Hungarian Academy of Sciences (MTA KFKI RMKI), H-1525 Budapest 114, POBox 49, Budapest, Hungary}
\newcommand{\korea}{Korea University, Seoul, 136-701, Korea}
\newcommand{\kurchatov}{Russian Research Center ``Kurchatov Institute", Moscow, Russia}
\newcommand{\kyoto}{Kyoto University, Kyoto 606-8502, Japan}
\newcommand{\labllr}{Laboratoire Leprince-Ringuet, Ecole Polytechnique, CNRS-IN2P3, Route de Saclay, F-91128, Palaiseau, France}
\newcommand{\lawllnl}{Lawrence Livermore National Laboratory, Livermore, California 94550, USA}
\newcommand{\losalamos}{Los Alamos National Laboratory, Los Alamos, New Mexico 87545, USA}
\newcommand{\lpc}{LPC, Universit{\'e} Blaise Pascal, CNRS-IN2P3, Clermont-Fd, 63177 Aubiere Cedex, France}
\newcommand{\lund}{Department of Physics, Lund University, Box 118, SE-221 00 Lund, Sweden}
\newcommand{\maryland}{University of Maryland, College Park, Maryland 20742, USA}
\newcommand{\mass}{Department of Physics, University of Massachusetts, Amherst, Massachusetts 01003-9337, USA }
\newcommand{\muenster}{Institut fur Kernphysik, University of Muenster, D-48149 Muenster, Germany}
\newcommand{\muhlenberg}{Muhlenberg College, Allentown, Pennsylvania 18104-5586, USA}
\newcommand{\myongji}{Myongji University, Yongin, Kyonggido 449-728, Korea}
\newcommand{\nagasaki}{Nagasaki Institute of Applied Science, Nagasaki-shi, Nagasaki 851-0193, Japan}
\newcommand{\newmex}{University of New Mexico, Albuquerque, New Mexico 87131, USA }
\newcommand{\nmsu}{New Mexico State University, Las Cruces, New Mexico 88003, USA}
\newcommand{\ornl}{Oak Ridge National Laboratory, Oak Ridge, Tennessee 37831, USA}
\newcommand{\orsay}{IPN-Orsay, Universite Paris Sud, CNRS-IN2P3, BP1, F-91406, Orsay, France}
\newcommand{\peking}{Peking University, Beijing, People's Republic of China}
\newcommand{\pnpi}{PNPI, Petersburg Nuclear Physics Institute, Gatchina, Leningrad region, 188300, Russia}
\newcommand{\riken}{RIKEN Nishina Center for Accelerator-Based Science, Wako, Saitama 351-0198, JAPAN}
\newcommand{\rikjrbrc}{RIKEN BNL Research Center, Brookhaven National Laboratory, Upton, New York 11973-5000, USA}
\newcommand{\rikkyo}{Physics Department, Rikkyo University, 3-34-1 Nishi-Ikebukuro, Toshima, Tokyo 171-8501, Japan}
\newcommand{\saispbstu}{Saint Petersburg State Polytechnic University, St. Petersburg, Russia}
\newcommand{\saopaulo}{Universidade de S{\~a}o Paulo, Instituto de F\'{\i}sica, Caixa Postal 66318, S{\~a}o Paulo CEP05315-970, Brazil}
\newcommand{\seoulnat}{System Electronics Laboratory, Seoul National University, Seoul, Korea}
\newcommand{\stonybrkc}{Chemistry Department, Stony Brook University, Stony Brook, SUNY, New York 11794-3400, USA}
\newcommand{\stonycrkp}{Department of Physics and Astronomy, Stony Brook University, SUNY, Stony Brook, New York 11794, USA}
\newcommand{\tenn}{University of Tennessee, Knoxville, Tennessee 37996, USA}
\newcommand{\titech}{Department of Physics, Tokyo Institute of Technology, Oh-okayama, Meguro, Tokyo 152-8551, Japan}
\newcommand{\tsukuba}{Institute of Physics, University of Tsukuba, Tsukuba, Ibaraki 305, Japan}
\newcommand{\vandy}{Vanderbilt University, Nashville, Tennessee 37235, USA}
\newcommand{\waseda}{Waseda University, Advanced Research Institute for Science and Engineering, 17 Kikui-cho, Shinjuku-ku, Tokyo 162-0044, Japan}
\newcommand{\weizmann}{Weizmann Institute, Rehovot 76100, Israel}
\newcommand{\yonsei}{Yonsei University, IPAP, Seoul 120-749, Korea}
\affiliation{\abilene}
\affiliation{\banaras}
\affiliation{\barc}
\affiliation{\bnlcoll}
\affiliation{\bnlphys}
\affiliation{\caucr}
\affiliation{\charlesczech}
\affiliation{\chonbuk}
\affiliation{\ciae}
\affiliation{\cns}
\affiliation{\colorado}
\affiliation{\columbia}
\affiliation{\czechtech}
\affiliation{\dapnia}
\affiliation{\debrecen}
\affiliation{\elte}
\affiliation{\ewha}
\affiliation{\fit}
\affiliation{\fsu}
\affiliation{\gsu}
\affiliation{\hiroshima}
\affiliation{\ihepprot}
\affiliation{\illuiuc}
\affiliation{\instpasczech}
\affiliation{\isu}
\affiliation{\jinrdubna}
\affiliation{\jyvaskyla}
\affiliation{\kek}
\affiliation{\kfki}
\affiliation{\korea}
\affiliation{\kurchatov}
\affiliation{\kyoto}
\affiliation{\labllr}
\affiliation{\lawllnl}
\affiliation{\losalamos}
\affiliation{\lpc}
\affiliation{\lund}
\affiliation{\maryland}
\affiliation{\mass}
\affiliation{\muenster}
\affiliation{\muhlenberg}
\affiliation{\myongji}
\affiliation{\nagasaki}
\affiliation{\newmex}
\affiliation{\nmsu}
\affiliation{\ornl}
\affiliation{\orsay}
\affiliation{\peking}
\affiliation{\pnpi}
\affiliation{\riken}
\affiliation{\rikjrbrc}
\affiliation{\rikkyo}
\affiliation{\saispbstu}
\affiliation{\saopaulo}
\affiliation{\seoulnat}
\affiliation{\stonybrkc}
\affiliation{\stonycrkp}
\affiliation{\tenn}
\affiliation{\titech}
\affiliation{\tsukuba}
\affiliation{\vandy}
\affiliation{\waseda}
\affiliation{\weizmann}
\affiliation{\yonsei}
\author{A.~Adare} \affiliation{\colorado}
\author{S.~Afanasiev} \affiliation{\jinrdubna}
\author{C.~Aidala} \affiliation{\mass}
\author{N.N.~Ajitanand} \affiliation{\stonybrkc}
\author{Y.~Akiba} \affiliation{\riken} \affiliation{\rikjrbrc}
\author{H.~Al-Bataineh} \affiliation{\nmsu}
\author{J.~Alexander} \affiliation{\stonybrkc}
\author{K.~Aoki} \affiliation{\kyoto} \affiliation{\riken}
\author{Y.~Aramaki} \affiliation{\cns}
\author{E.T.~Atomssa} \affiliation{\labllr}
% ...
\collaboration{PHENIX Collaboration} \noaffiliation

\date{\today}


\begin{abstract}

Differential measurements of the elliptic ($v_2$) and hexadecapole ($v_4$) 
Fourier flow coefficients are reported for charged hadrons as a function of 
transverse momentum ($p_T$) and collision centrality or number of participant 
nucleons ($N_{\rm part}$) for Au+Au collisions at $\sqrt{s_{NN}}~=~200$ GeV.  
The $v_{2,4}$ measurements at pseudorapidity $\left| \eta \right| \le 0.35$, 
obtained with four separate reaction-plane detectors positioned in the range 
$1.0 < \left| \eta \right| < 3.9$ show good agreement, indicating the absence 
of significant $\Delta\eta$-dependent nonflow correlations.  Sizable values 
for ${v_{4}(p_T)}$ are observed with a ratio $\frac{v_4(p_T, N_{\rm part})} 
{v_2^2(p_T, N_{\rm part})} \approx 0.8$ for $50 \alt N_{\rm part} \alt 200$, 
which is compatible with the combined effects of a finite viscosity and 
initial eccentricity fluctuations.  For $N_{\rm part} \agt 200$ this ratio 
increases up to $1.7$ in the most central collisions.

\end{abstract}


\maketitle


The discovery of large azimuthal anisotropy at the Relativistic Heavy Ion 
Collider (RHIC) is a key piece of evidence for the creation of dense partonic 
matter in ultra relativistic nucleus-nucleus collisions 
\cite{Adcox:2004mh,Adams:2005dq}.  With sufficiently strong interactions, the 
medium in the collision zone can be expected to locally equilibrate and 
exhibit hydrodynamically driven flow 
\cite{Ollitrault:1992bk,Heinz:2001xi,Shuryak:2008eq}.  The momentum anisotropy 
results from an initial ``almond-shaped" collision zone produced in noncentral 
collisions \cite{Ollitrault:1992bk,Heinz:2001xi}.  It is now routinely 
characterized, at midrapidity, by the even order Fourier coefficients 
$v_n = \mean{e^{in(\phi_p - \Phi_{\rm RP})}}, n=2,4,...,$ where $\phi_p$ is the 
azimuthal angle of an emitted particle, $\Phi_{\rm RP}$ is the azimuth of the 
reaction plane and the brackets denote averaging over particles and events.

At the highest RHIC collision energy of $\sqrt{s_{NN}}=200$\,GeV, differential 
elliptic flow measurements $v_2(p_T)$ (for transverse momentum $p_T{\,\alt\,} 
2.5$\,GeV/$c$) and $v_2(N_{\rm part})$ have been measured for a broad range of 
centralities or number of participants $N_{\rm part}$.  These data are found to 
be in accord with calculations that model an essentially locally equilibrated 
quark gluon plasma (QGP) having little or no viscosity 
\cite{Heinz:2001xi,Adare:2006nq,Romatschke:2007mq,Xu:2007jv}.  Quark number 
scaling of elliptic flow data (suggestive of partonic degrees of freedom in 
the collision zone) is observed for a broad range of particle species, 
collision centralities and transverse kinetic energy 
\cite{Adare:2006ti,Lacey:2006pn}.  Small violations of the scaling of 
$v_2(N_{\rm part})$ with the initial eccentricity of the collision zone 
$\varepsilon$, suggest a strongly-coupled low-viscosity plasma 
($4\pi\frac{\eta}{s} \sim 1 - 2$ for the ratio of viscosity $\eta$ to entropy 
density $s$) in energetic Au+Au collisions \cite{Lacey:2006bc, 
Drescher:2007cd,Chaudhuri:2009ud}.  Nonetheless, the degree to which the QGP is 
thermalized \cite{Borghini:2005kd}, and whether it is strongly or weakly 
coupled \cite{Shuryak:2008eq,Asakawa:2006tc}, is still being debated.

Recent theoretical studies indicate that the hexadecapole flow harmonic $v_4$ 
is a more sensitive constraint on the magnitude of $\frac{\eta}{s}$ and the 
freeze-out dynamics \cite{Greco:2008fs}, and the ratio $\frac{v_4}{(v_2)^2}$ 
can indicate whether full local equilibrium is achieved in the QGP 
\cite{Bhalerao:2005mm}.  The role of fluctuations and so-called ``nonflow'' 
correlations is important for such measurements.  It is well established that 
initial eccentricity fluctuations significantly influence the magnitudes of 
$v_{2,4}$ \cite{Miller:2003kd,Hama:2007dq,Alver:2006wh,Lacey:2009xx,Gombeaud:2009ye}.  
However, the precise role of nonflow, which leads to a systematic error 
in the determination of $v_{2,4}$ is less clear.  Non-flow correlations among 
produced particles may arise from jets, whose influence is found to vary with 
pseudorapidity $\eta$ and $p_T$ \cite{Jia:2006sb}.  This provides a tool to 
evaluate how jets influence the measurements presented here.

We report precise measurements of charged hadron $v_2$ and $v_4$ in Au+Au 
collisions at $\sqrt{s_{NN}}=200$\,GeV.  The measurements were performed in the 
two PHENIX central arms ($\left| \eta \right| \le 0.35$) with respect to event 
planes obtained from four separate reaction-plane detectors in the range $1.0 
< \left| \eta \right| < 3.9$.  Multiple event planes allow a search for 
possible $\Delta\eta$-dependent nonflow contributions that would influence the 
magnitude of $v_{2,4}$, which may be crucial for reliable extraction of 
transport coefficients.

The results reported here are derived from $\sim 3.6 \times 10^9$ minimum-bias 
Au+Au events obtained at $\sqrt{s_{NN}}=$ 200~GeV with the PHENIX 
detector~\cite{Adcox:2003zm} during the 2007 running period.  The event 
centrality was determined via cuts on the analog response of the Beam-Beam 
Counters (BBC).  For each centrality selection, the number of participant 
nucleons $N_{\rm part}$, was estimated via a Glauber model Monte-Carlo 
simulation \cite{Alver:2008aq}.  The drift chambers and two layers of 
multi-wire proportional chambers with pad readout (PC1 and PC3) were used for 
charged particle tracking and momentum reconstruction with azimuthal coverage 
$\Delta\varphi=\pi/2$ in the central region ($|\eta|\leq 0.35$).  
Tracks were required to have $E/p_T > 0.1$ and a confirmation hit within
a $2 \sigma$ matching window in PC3 and the Electromagnetic
Calorimeters (EMCal) ($E$ denotes the energy deposited in the EMCal).
This minimized albedo, conversions and weak decay products.

\begin{figure}[tb]
\includegraphics[width=1.0\linewidth]{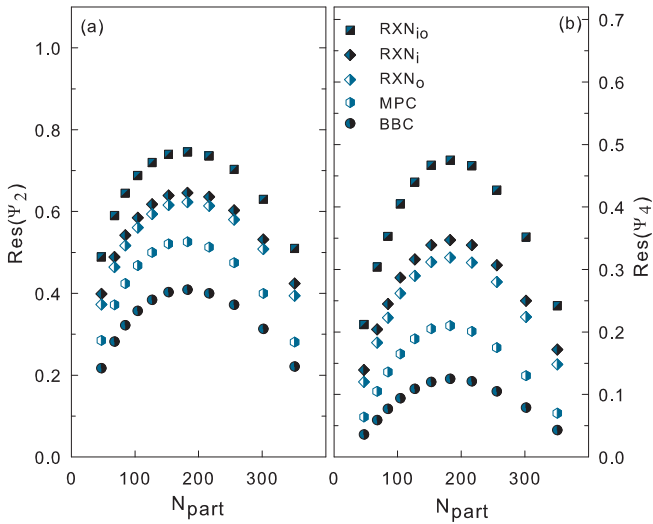}
\caption{(color online) Event-plane resolution factors vs.  $N_{\rm part}$ 
for 
$v_2$ (a) and $v_4$ (b) 
measurements for the indicated event planes.  
}
\label{Fig1}
\end{figure}

The event-plane method~\cite{Afanasiev:2009wq} was used to correlate the 
azimuthal angles $\phi_p$ of the charged tracks in the PHENIX central arms 
($\left| \eta \right| \le 0.35$) with the azimuth of the estimated second 
order event plane $\Phi_2$, determined via hits in the two BBCs and Muon 
Piston Calorimeters (MPCs), and the two inner (i), outer (o) and combined (io) 
rings of newly installed Reaction-Plane Detectors (RXN).  The two RXNs are 
situated at $|z|~=$~38--40~cm of the nominal crossing point and their inner and 
outer rings are comprised of twelve plastic scintillators ($\Delta \phi = 
\pi/6$ for each).  The MPCs are PbWO$_4$ based electromagnetic calorimeters 
with 2$\pi$ azimuthal acceptance.  The respective $\eta$ coverage for these 
event-plane detector pairs are $3.1 < \left| \eta_{_{\rm BBC}} \right| < 3.9 
$, $3.1 < \left| \eta_{_{\rm MPC}} \right| \alt 3.7 $, $1.5 < \left| 
\eta_{{\rm RXN_i}} \right| < 2.8$ and $1.0 < \left| \eta_{{\rm RXN_o}} \right| 
< 1.5$.  For a given pair the detector, which is located at positive (negative) 
$\eta$, is designated North (N) (South (S)).

Charge-averaged values for the second and fourth flow harmonics
were evaluated separately for each estimated event plane $i$ as:
\begin{equation}
v_{2k}^i = \frac{\left\langle \cos(2k(\phi_p -\Phi_2^i))\right\rangle} 
{{\rm Res}(\Psi_{2k}^i)}\;\; k =1,2 \;,
\label{v2v4}
\end{equation} 
where the denominator represents a resolution factor that corrects for the 
difference between the true azimuth $\Phi_{\rm RP}$ and the $2^{{\rm nd}}$ 
order estimate $\Phi_2^i$ of the event plane.  This estimate was obtained 
from the combined sub-events (North and South) for each detector pair.  
Resolution factors were evaluated via the three-sub-events method 
\cite{Poskanzer:1998yz,Afanasiev:2009wq}:
\begin{eqnarray}
 {\rm Res}(\Psi_{2k}^i) 
 = \sqrt{ \frac{\la \cos(2k(\Phi_2^{i}-\Phi_2^{l})) \ra
 \la \cos(2k(\Phi_2^{i}-\Phi_2^{m})) \ra } 
 {\la \cos(2k(\Phi_2^{l}-\Phi_2^{m})) \ra}},
\label{3subEv} 
\end{eqnarray}
where ${i,l\; {\rm and}\; m}$ indicate event and subevent planes with 
disparate $\eta$ values (eg., $i$ = RXN$_{{\rm io}}$, $l$ = MPC$_{N}$, and $m$ 
= BBC$_{{\rm S}}$).  An advantage of this procedure is that, for any given 
centrality, it allows several independent estimates of Res($\Psi_{2,4}$) for 
each event plane.  In turn, such estimates allow an evaluation of the 
systematic errors for Res($\Psi_{2,4}^i$).  It is noteworthy that estimates for 
these correction factors were also obtained (for $k = 1\; {\rm and}\; 2$) via 
the two-sub-events method \cite{Poskanzer:1998yz,Afanasiev:2009wq}, which is 
regularly used for elliptic flow analysis.  For RXN the difference between 
both methods is small for $v_2$ i.e, $\sim 1$\% for mid-central collisions and 
$\sim 5$\% for the most central and peripheral collisions.  For $v_4$, it is 
$\sim 2$\% for mid-central collisions and grows to $\sim 7$\% and 20\% in the 
most peripheral and central collisions respectively.

\begin{figure}[t]
\includegraphics[width=1.0\linewidth]{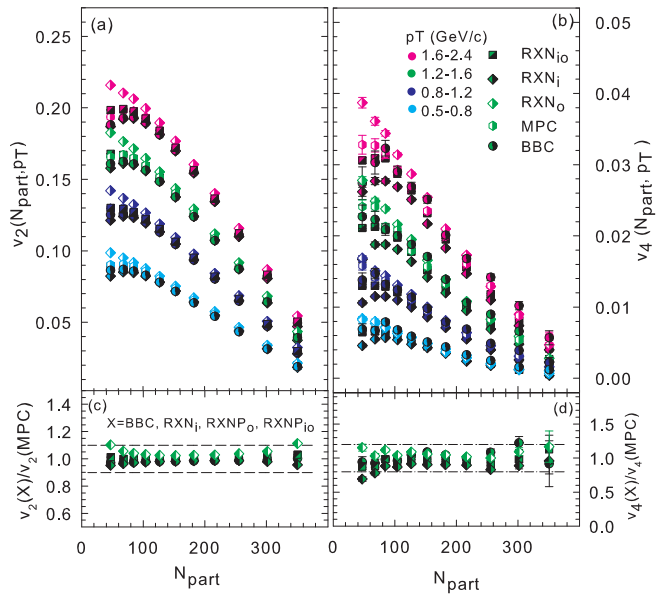}
\caption{(color online) Comparison of $v_2$ vs.  $N_{\rm part}$ (a) and $v_4$ 
vs.  $N_{\rm part}$ (b) for charged hadrons obtained with several reaction 
plane detectors for the $p_T$ selections indicated.  Ratios for the $p_T$ range 
1.2--1.6 GeV/$c$ are shown in (c) and (d); the curves indicate $\pm 10$\% and 
$\pm 20$\% systematic error bands.  }
\label{Fig2}
\end{figure}
    
Figure~\ref{Fig1} shows the centrality dependence of $\la {\rm Res}(\Psi_{2}) 
\ra$ and $\la {\rm Res}(\Psi_{4}) \ra$ for several event planes.  Similar 
maxima are observed for $N_{\rm part} \approx 200$ with a falloff at lower and 
higher $N_{\rm part}$.  Measurements with the ${\rm RXN}_{{\rm io}}$ event 
plane benefit from about a factor of two (five) improvement in the resolution 
for $v_2$ ($v_4$) compared to prior PHENIX measurements with the BBC event 
plane~\cite{Afanasiev:2009wq}.

The systematic errors associated with the ${\rm RXN}_{{\rm io}}$ resolution 
factors for $v_2$ ($v_4$) are estimated to be less than 2\% (6\%) for 
mid-central collisions but increase to about 3\% (10\%) in the most central 
and peripheral collisions.  Similar estimates were obtained for the ${\rm 
RXN}_{{\rm i}}$ and ${\rm RXN}_{{\rm o}}$ event planes.  On average, those for 
the BBC and the MPC event planes are about a factor of two larger.  Other 
sources, such as track cuts, are estimated to range from $\sim 1-2$\% (3-4\%) 
for $p_T \agt 0.5$~GeV/$c$ to $\sim$5\% (10\%) for the lowest $p_T$ values.
 
Figures \ref{Fig2}(a) and (b) compare the double differential flow 
coefficients $v_{2,4}(p_T, N_{\rm part})$ for event-plane detectors 
spanning 
the range $1.0 < \left| \eta \right| < 3.9$.  Within systematic errors, they 
agree to better than $\sim$ 5\% (10\%) for $v_2$ ($v_4$) in mid-central 
collisions and approximately 10\% (20\%) in central and peripheral events 
(cf., ratios in Figs.~\ref{Fig2}(c) and (d)) independent of $p_T$.  This 
agreement indicates a reliable measurement free of significant $\Delta\eta$- 
and $p_T$-dependent nonflow contributions (for $p_T \alt 3$ GeV/$c$), which 
would affect $v_2$ and $v_4$ (very little influence is expected from 
a possible $\Delta\eta$-independent long-range correlation \cite{Alver:2010rt}). 
Non-flow correlations, such as from dijets, would 
lead to a difference in the $v_2$ ($v_4$) values obtained with event planes 
determined at different rapidity gaps ($\Delta\eta$) with respect to the 
central arms \cite{Jia:2006sb}.  
In the following we utilize the RXN$_{\rm io}$ 
event plane due to its good resolution.  The associated systematic error for 
$v_2$ ($v_4$) is estimated to be $\approx 3$\% (8\%) for mid-central 
collisions and increase to about 7\% (15\%) in the most peripheral and 
central collisions.

\begin{figure}[t]
\includegraphics[width=1.0\linewidth]{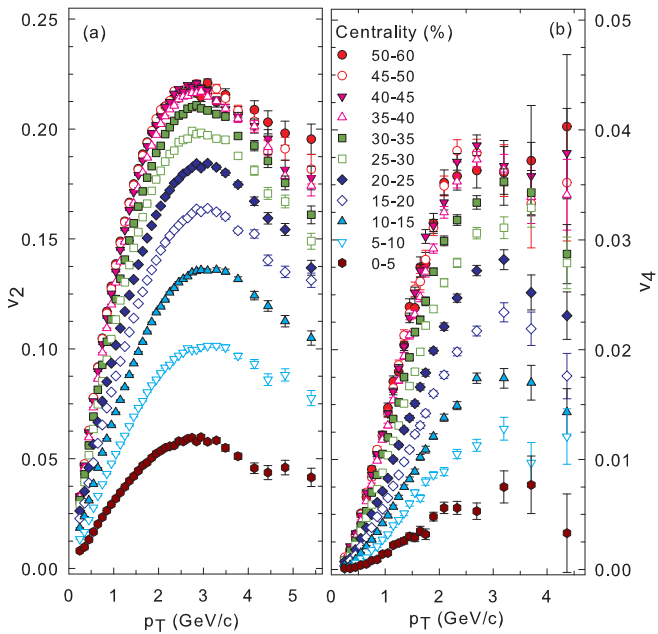} 
\caption{(color online) $p_T$ dependence of $v_2$ (a) and $v_4$ (b) for charged 
hadrons for several centrality selections as indicated.  The error bars only 
indicate statistical errors.
}
\label{Fig3}
\end{figure}

\begin{figure}[t]
\includegraphics[width=1.0\linewidth]{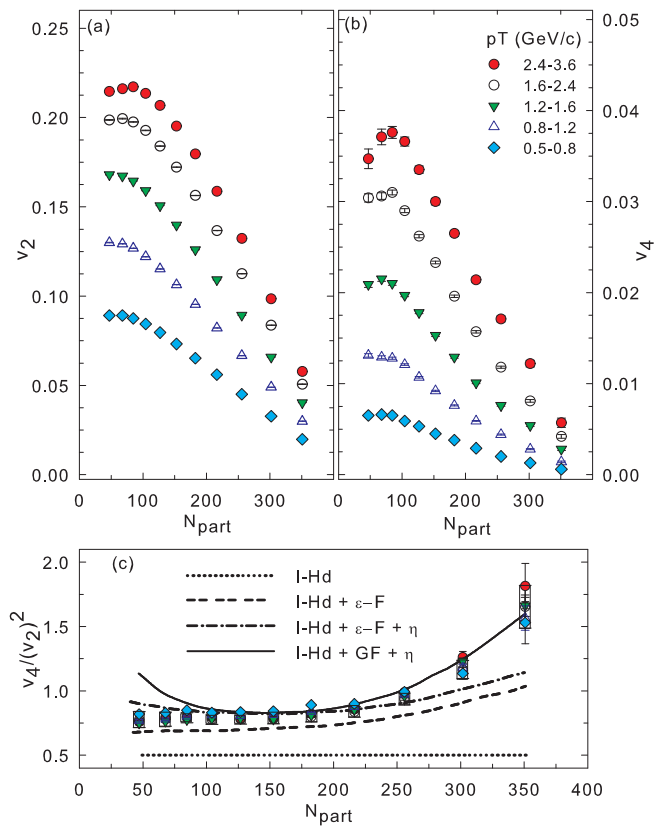} 
\caption{(color online) $v_2$ vs.  $N_{\rm part}$ (a) and $v_4$ vs.  $N_{\rm 
part}$ (b) for charged hadrons for several $p_T$ selections as indicated.  
Panel (c) shows the ratio $v_4/(v_2)^2$ vs.  $N_{\rm part}$ for the same $p_T$ 
selections.  The open boxes indicate systematic errors for the selection 
$1.6 < p_T < 2.4$ GeV/$c$.  The curves show calculated results for ideal 
hydrodynamics (I-Hd), I-Hd + eccentricity fluctuations ($\varepsilon$-F), 
I-Hd + $\varepsilon$-F + viscosity ($\eta$), and I-Hd + $\eta$ + Gaussian 
Fluctations (GF) (see~\cite{Gombeaud:2009ye}).  }
\label{Fig4} 
\end{figure}

Figures~\ref{Fig3} and~\ref{Fig4} summarize the results for elliptic and 
hexadecapole flow.  The magnitude and trends agree well with those reported 
earlier \cite{Adcox:2004mh,Adams:2005dq}.  However they now benefit from a 
factor of five increase in statistics, as well as improved precision ($\sim 
2$) in the event plane.  Figures \ref{Fig3}(a) and (b) compare the 
measured charged hadron differential $v_2(p_T)$ and $v_4(p_T)$, as a function 
of centrality.  In contrast to the approximately linear dependence observed in 
Fig.~\ref{Fig3}(a) for $p_T \alt 1.5 $~GeV/$c$, the $v_4$ data exhibit a non 
linear dependence on $p_T$ compatible with the prediction from hydrodynamics 
that $v_4 \propto v_2^2$ \cite{Kolb:2003zi}.  The large increase ($\sim \times 6$) 
from central to peripheral collisions, reflects the expected increase due to the 
change in initial eccentricity from central to peripheral 
events \cite{Bhalerao:2005mm,Broniowski:2007ft}.

Figure~\ref{Fig4} compares the $v_2(N_{\rm part})$ (a) and $v_4(N_{\rm part})$ 
(b) for several $p_T$ selections as indicated.  The $N_{\rm part}$ values are 
mean values evaluated for the centrality selections indicated in 
Fig.~\ref{Fig3}.  Here, the data trends in (a) and (b) are strikingly similar 
albeit with a much smaller magnitude in (b).  The magnitude and trends with 
$p_T$ and $N_{\rm part}$ in Figs.~\ref{Fig4}(a) and (b) follow expectations 
for a hydrodynamically expanding low viscosity fluid 
\cite{Lacey:2006bc,Romatschke:2007mq,Xu:2007jv,Shuryak:2008eq,Drescher:2007cd, 
Chaudhuri:2009ud}.

The ratio $\frac{v_4}{(v_2)^2}$ is shown as a function of $N_{\rm part}$ in 
Fig.~\ref{Fig4}(c) for the same $p_T$ selections used in (a) and (b); 
systematic errors are $\approx 4-5$\% for mid-central collisions and increase 
to 8--10\% for central and peripheral collisions.  Within errors, these data 
indicate that the magnitude of $\frac{v_4}{(v_2)^2}$ is essentially 
independent of $p_T$ for the range $0.5 - 3.6$ GeV/$c$, {\em i.e.} extending 
beyond the maxima in Fig.  \ref{Fig3}(a).  An approximately constant ratio of 
value $\frac{v_4(p_T, N_{\rm part})}{v_2^2(p_T, N_{\rm part})} \approx 0.8$ is 
observed for $50 \alt N_{\rm part} \alt 200$, which is larger than the ratio 
$\approx 0.5$ for ideal hydrodynamics in the model of \cite{Gombeaud:2009ye}.
The inclusion of eccentricity fluctuations in this model, cause this ratio 
to exceed 0.5 as shown by the dashed curve (from~\cite{Gombeaud:2009ye}) 
in Fig.~\ref{Fig4}(c).  
Viscosity from the hadron gas phase, in addition to a small value in the quark 
gluon plasma ($4\pi\frac{\eta}{s} \sim 2$) \cite{Drescher:2007cd}, results in 
a further increase of this ratio as indicated by the dashed-dot curve 
\cite{Gombeaud:2009ye}.

Our $\frac{v_4(p_T, N_{\rm part})}{v_2^2(p_T, N_{\rm part})}$ ratio is smaller 
than the centrality-averaged value of 1.2 reported by STAR 
\cite{Adams:2003zg}.  Part of this difference can be understood by averaging 
over our measured centrality range (0-60\%) yielding the value $\approx 1.0$.  
Comparison to STAR results~\cite{Gombeaud:2009ye} shows a 10\% 
discrepancy for mid-central collisions, possibly reflecting differences in 
the methods used to estimate Res($\Psi_{4}$).

In more central collisions where $N_{\rm part} \agt 200$, $\frac{v4}{v_2^2}$ 
increases rapidly.  Adding eccentricity fluctuations to ideal hydrodynamics 
causes a similar trend, indicated by the dashed curve in Fig.~\ref{Fig4}(c).  
Central collisions are the most sensitive because the eccentricity decreases 
as the overlap region becomes more symmetric.  In order to reproduce the 
central data, the authors of \cite{Gombeaud:2009ye} introduced additional 
fluctuations shown as the solid line in Fig.~\ref{Fig4}(c), though the source 
of these fluctuations is as yet unspecified.


In summary, we have presented differential measurements of $v_4$ and $v_2$ for 
charged hadrons obtained with four reaction-plane detectors at different 
$\Delta\eta$ with respect to the PHENIX central arms.  There are no significant 
$\Delta\eta$- and $p_T$-dependent nonflow contributions for $p_T \alt 3$ 
GeV/$c$ in the centrality ranges of our study.  Consequently there are no 
significant systematic errors from jets on the event-plane determinations 
or 
values of $v_2$ and $v_4$.  The ratio $\frac{v_4(p_T, N_{\rm part})} 
{v_2^2(p_T, N_{\rm part})} \approx 0.8$ for $50 \alt N_{\rm part} \alt 200$ is 
essentially independent of $p_T$, consistent with the effects of finite 
viscosity and eccentricity fluctuations.  For $N_{\rm part} \agt 200$ the ratio 
increases up to $1.7$ in the most central collisions.  The precision of these 
data provide stringent constraints for further theoretical modeling and more 
detailed extractions of the transport properties of hot and dense partonic 
matter.



We thank the staff of the Collider-Accelerator and 
Physics Departments at BNL for their vital contributions.  
We acknowledge support from 
the Office of Nuclear Physics in DOE Office of Science and NSF (U.S.A.), 
MEXT and JSPS (Japan), 
CNPq and FAPESP (Brazil), 
NSFC (China), 
MSMT (Czech Republic),
IN2P3/CNRS and CEA (France), 
BMBF, DAAD, and AvH (Germany), 
OTKA (Hungary), 
DAE and DST (India), 
ISF (Israel), 
NRF (Korea), 
MES, RAS, and FAAE (Russia),
VR and KAW (Sweden), 
U.S.  CRDF for the FSU, 
US-Hungary Fulbright, 
and US-Israel BSF.



\end{document}